\documentclass[aps,preprint,twocolumn,superscriptaddress,10pt]{revtex4}%
\usepackage{amsfonts}
\usepackage{amsmath}
\usepackage{amssymb}
\usepackage{graphicx}%
\providecommand{\U}[1]{\protect \rule{.1in}{.1in}}
\usepackage[colorlinks, citecolor=blue]{hyperref}

\begin{document}

\title[ ]{The credibility of Rydberg atom based digital communication over a continuously tunable radio-frequency carrier}
\author{Zhenfei Song\footnote{E-mail:songzf@nim.ac.cn}}
\affiliation{National Institute of Metrology, Beijing 100029, People's Republic of China}
\author{Wanfeng Zhang}
\affiliation{National Institute of Metrology, Beijing 100029, People's Republic of China}
\author{Hongping Liu\footnote{E-mail:liuhongping@wipm.ac.cn}}
\affiliation{State Key Laboratory of Magnetic Resonance and Atomic and Molecular Physics,
Wuhan Institute of Physics and Mathematics, Chinese Academy of Sciences, Wuhan
430071, People's Republic of China}
\author{Xiaochi Liu}
\affiliation{National Institute of Metrology, Beijing 100029, People's Republic of China}
\author{Haiyang Zou}
\affiliation{National Institute of Metrology, Beijing 100029, People's Republic of China}
\affiliation{Southeast University, Nanjing 210096, People's Republic of China}
\author{Jie Zhang}
\affiliation{National Institute of Metrology, Beijing 100029, People's Republic of China}
\affiliation{China Jiliang University, Hangzhou 310018, People's Republic of China}
\author{Jifeng Qu}
\affiliation{National Institute of Metrology, Beijing 100029, People's Republic of China}
\pacs{PACS number}

\begin{abstract}
High-sensitive measurement of radio-frequency (RF) electric field is available via the electromagnetically induced transparency (EIT) effect of Rydberg atom at
room-temperature, which has been developed to be a promising atomic RF receiver. In this Letter, we investigate the credibility of the digital communication via this quantum-based antenna over the entire continuously tunable RF-carrier.
Our experiment shows that digital communication at a rate of 500 kbps performs reliably within a tunable bandwidth of 200 MHz at carrier 10.22 GHz
 and a bit error rate (BER) appears out of this range, for example, the BER runs up to 15{\%} at RF-detuning $\pm150$ MHz. 
In the measurement, the time-variant RF field is retrieved by detecting the density of the probe laser at the center frequency of RF-induced symmetric or asymmetric Autler-Townes splitting in EIT.
Prior to the digital test, we have studied the RF-receiving quality versus the physical ambiance and found that a choice of linear gain response to the RF-amplitude can suppress the signal distortion and the modulating signal is able to be decoded as fast as up to 500 kHz in the tunable bandwidth. Our checkout consolidates the physical foundation for a reliable communication and spectrum sensing over the broadband RF-carrier.

\end{abstract}
\date{\today}
\maketitle
\volumeyear{ }
\volumenumber{ }
\issuenumber{ }
\eid{ }
\received{Received text}{}

\revised{Revised text}{}

\accepted{Accepted text}{}

\published{Published text}{}

\startpage{1}
\endpage{ }

Over the last decades, a wide variety of experimental quantum communications and processing devices have been developed for fundamental demonstrations in laboratories, which validates the feasibility of practical applications in communications and information related fields. Applications in areas like
quantum sensors \cite{01Kitching2011,02Degen2017a} are partially commercially available. Atom-based quantum techniques are emerging as a completely new and promising paradigm for advanced communications.

Accurate radio-frequency (RF) electromagnetic field sensing in free-space plays a fundamental role in wireless communication, and Rydberg atoms are remarkable
quantum sensors for RF electric (E-) field measurements \cite{03Sedlacek2012,04Fan2015a}.
Applied RF E-field induces strong ac-stark coupling between Rydberg states, resulting in an Autler-Townes (AT) splitting of a ladder-type electromagnetic induced transparency (EIT) \cite{06Petrosyan2011}, which can convert the measurement of RF field into optical frequency determination \cite{03Sedlacek2012}. 
Comparing with conventional methods, this quantum-optical method has advantages of high sensitivity with a predicted shot noise limit of pV$\cdot$cm$^{ - 1}\cdot$Hz$^{ - 1 / 2}$ \cite{07Kumar2017}, high accuracy with an expected measurement uncertainty of $0.5{\%}$ \cite{03Sedlacek2012,08Holloway2017},
ultra-broadband measurement covering from $\sim$100 MHz to THz \cite{09Holloway2014e,10Jiao2017a}, and atom-based self-calibration. 
This method is promising to become a new generation of RF E-field measurement standard \cite{11Holloway2017c}.

In particular, some proof-of-concept work on wireless communication and remote sensing using Rydberg atoms have been presented recently \cite{17Meyer2018,Deb2018,Jiao2018}. 
The time-variant E-field signal in free-space can be captured by measuring the transmission of a probe laser in a condition of a Rydberg EIT. Owing to unique advantages of free-space RF field sensing, the quantum receiver has great significance compared with conventional electronics-based receivers, including but not limited to the
weak signal, long-distance communication in free space or via a fiber link.
All the principle experiments of communication were performed over carrier of an optimized resonant frequency of Rydberg states \cite{17Meyer2018,Deb2018,Jiao2018}.

Nowadays the wireless communication spectrum resources are increasingly scarce, and cognitive radio technology is being used to provide a solution for more efficient utilization of the radio spectrum \cite{Haykin07}. As the very essence of cognitive radio, continuous turnable RF receiving which covers an entire frequency band has become increasingly important for accessing spare subbands of the radio spectrum that are underutilized.
In addition, the broadband communication in higher frequency bands can provide high speed, high capacity data transfer, these are the most exciting features of the fifth generation (5G) communication \cite{Boccardi2014}.
In a word, rather than retrieving the time-variant signals over a specific carrier frequency, we hope to make extensive use of the frequency band near the carrier determined by the transition between the neighboring Rydberg states. It requires us to have this Rydberg atom based quantum-optical radio communication features over the whole RF-carrier near the Rydberg resonance in hand.

In this Letter, we first study the spectral features of the modulating signal transferring over different carrier frequency detunings relative
to the resonant transition, such as the detecting efficiency and linearity on the modulation frequency and carrier power, as well as  the signal-to-noise ratio (SNR) on the physical ambiance over the RF detuning. 
After that we further qualitatively research the RF signal transfer properties over the carrier frequency domain by testing digital communication using a benchmark of pseudo-random binary sequence (PRBS) signal.
The experiment shows that the digital communication at a rate of 500 kbps exhibits a reliable performance within the tunable bandwidth of 200 MHz over 10.22 GHz carrier and a bit error rate (BER) of  15{\%} appears at RF-detuning of $\pm150$ MHz.

\begin{figure}[ptbh]
\centerline{\includegraphics[width=3.30in]{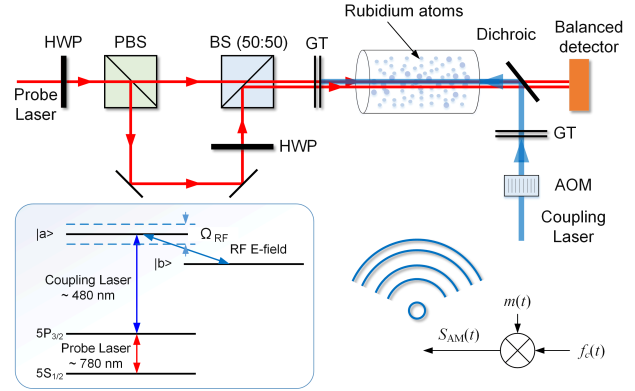}}
\caption{Experimental setup of the Rydberg atom based RF-receiver with a differential detection. The coupling and probe lasers counter-propagate through the rubidium vapor cell, forming a ladder-type EIT. A RF E-field couples the two Rydberg states labeled as $\vert a>$ and $\vert b>$, resulting in an Aulter-Townes splitting subsequently.
The inset is a relevant energy level of the Rb atoms involved in the work. HWP: half-wavelength plate; PBS: polarizing beam splitter; BS: beam splitter; GT: Glan-Taylor prism; AOM: acoustic optic modulator.}
\label{fig1}
\end{figure}

The Rydberg atom EIT-based RF-receiving is schemed in Fig. \ref{fig1} with the related energy levels of $^{87}$Rb shown in the inset. It is similar to the reference works \cite{17Meyer2018,Deb2018,Jiao2018} except the (de)coding part of the signal over the RF carrier and the differential detection. As shown in the inset, the long-lived  ground state $5S_{1/2}$ and high Rydberg state $\vert a>$ are coupled to a short-lived state $5P_{3/2}$ by a weak probe laser of 780 nm and an intense coupling laser of 480 nm, respectively, forming a typical ladder-type EIT system. The probe and coupling lasers are schemed to counter-propagate in rubidium vapor gas with Doppler broadening effect excluded maximally \cite{07Kumar2017}.
The atomic vapor cell has a 8 mm cubic size and operates at room-temperature (30$\pm$0.5${^{\circ}}$C). The lasers with linewidth less than $\sim$100 kHz are focused to full-width-half-maxima (FWHMs) of 120 \textit{$\mu$}m and 200 \textit{$\mu$}m, respectively. With the probe laser locked to the resonant transition from $5S_{1/2}$ to $5P_{3/2}$, we can record a clear EIT signal by scanning the coupling laser.
Extremely linearly polarized lasers produced by Glan-Taylor prisms are employed to achieve a high-contrast EIT signal. In addition, the differential detection technique has been employed to remove the futile background signal and to improve the SNR, rather than the usual configuration where  elliptically polarized lasers are used \cite{18Lin2017}. In our setup, a pair of HWP and PBS is adopted to balance the transmissive and reflective laser intensities, while another pair of HWP and BS used to collimate and co-polarize the 90$^\circ$-rotated beam, serving as a reference signal for the balance detector.

If we apply a RF field coupling another Rydberg state $\vert b>$ to $\vert a>$, a further spectral splitting will occur within the EIT peak subsequently.
Due to the large dipole moment of Rydberg states, even a weak RF E-field can also induce a  strong coupling, leading to a remarkable ramification, known as AT-splitting, which makes it capable to serve as a RF receiver in wireless communication.
The carrier of 10.22 GHz is generated by a microwave signal generator (Agilent N5183A) which is amplitude-modulated externally by a kHz
RF generator. The modulated signal at the carrier is radiated towards the Rydberg atoms in the cell by an X-band horn antenna with the far-field gain of 15.1 dBi.

\begin{figure}[ptbh]
\centerline{\includegraphics[width=3.3in]{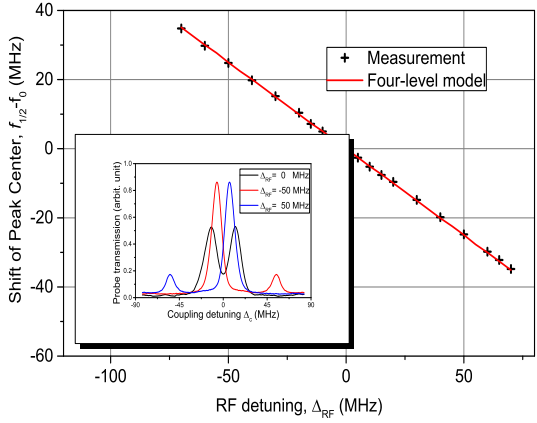}}
\caption{The linear dependence of the frequency shift of AT-splitting center, $f_{1/2}-f_0$, on the RF detuning $\Delta_{RF}$.
A pair of typical asymmetric AT-splitting spectra is shown in the inset at RF detuning $\pm 50$ MHz relative to the energy difference between Rydberg states 59 $D_{5 / 2}$ and 60 $P_{3 / 2}$ (10.22 GHz).
}
\label{fig2}
\end{figure}

To make a full use of the bandwidth of the RF carrier of the Rydberg atom based receiver, we have a look at the spectral feature of the probe beam at different RF detunings.
When the applied RF field is frequency-detuned from the resonant transition between
two Rydberg states, the observed spectroscopy of AT-splitting changes as well, becoming asymmetric as shown in the inset of Fig. \ref{fig2}, for example, at RF field detunings at $\pm50$ MHz near a resonant frequency of 10.22 GHz for Rydberg transition from 59 $D_{5/2}$ to 60 $P_{3/2}$.
The center frequency of the asymmetric AT-splitting separation, denoted by $f_{1/2}$, varies with RF-detuning $\Delta_{RF}$ as well.
The amount of the variation of $f_{1/2}$ with respect to the resonant coupling laser frequency at zero RF-detuning, labeled as $f_0$, satisfies the
relation $f_{1/2}-f_0=-\frac{1}{2}\Delta_{RF}$, as numerically shown in Fig. \ref{fig2}.
This relation is supported by a four-level model simulation \cite{08Holloway2017,19SIBALIC2017}. In the succedent experiment, an AOM together with an external driven source is implemented to accurately shift the coupling laser frequency according to a specific RF detuning, and we will monitor the probe beam intensity at  $f_{1/2}$  for characterizing RF-field variation where the photodiode has the most sensitive response to the applied RF-field. This simple scheme supplies a faster transfer rate than the method of encoding information into the AT-splitting width \cite{Jiao2018}.

\begin{figure}[ptbh]
\centerline{\includegraphics[width=3.3in]{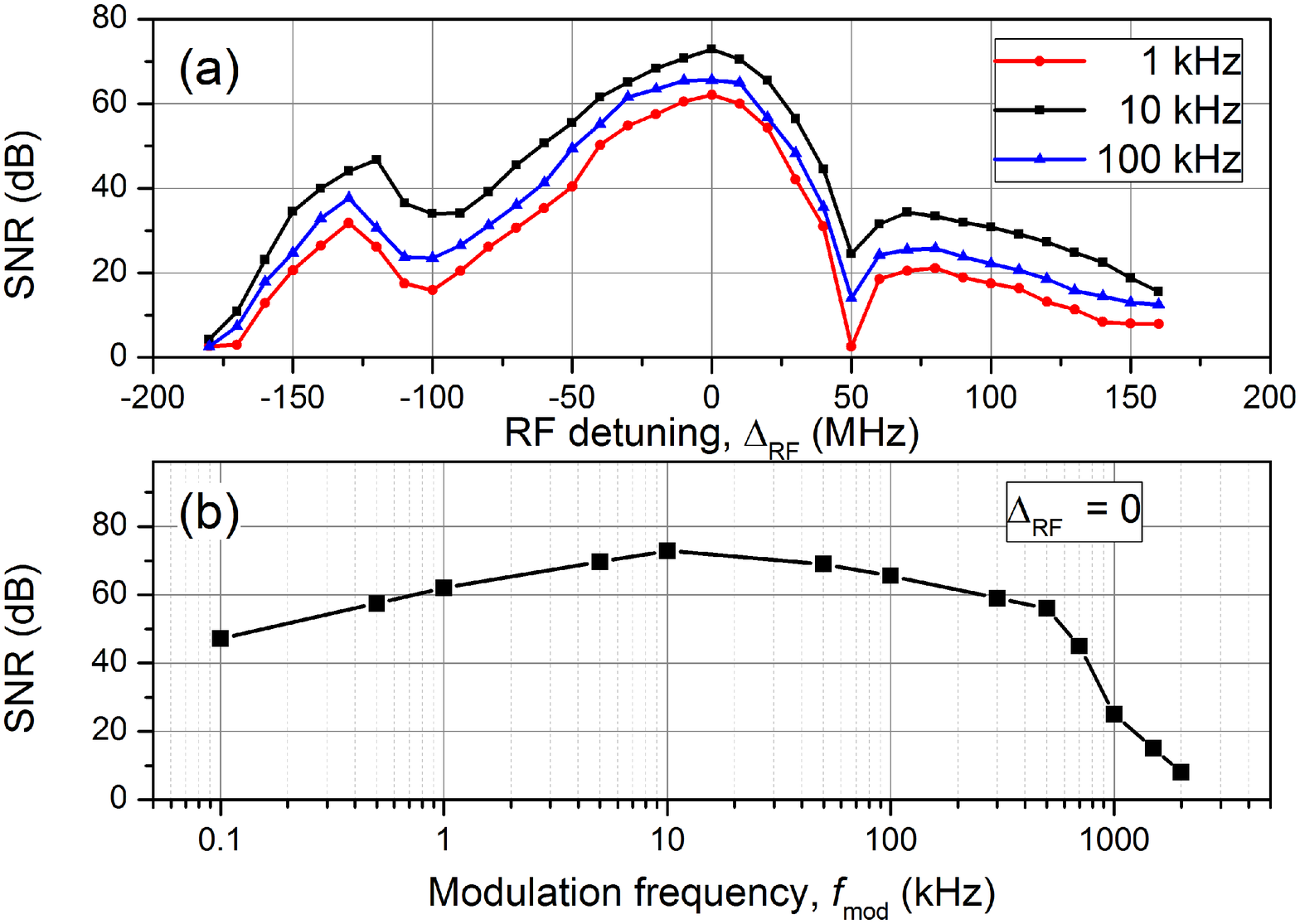}}
\caption{The SNR of the differential demodulated signal on the balanced detector versus the RF-detunings at several modulation frequencies $\emph{f}_{mod}$ (a) and its dependence on the modulation frequency at finer steps of $\emph{f}_{mod}$ (b). The SNR has an considerable values in a carrier bandwidth up to $200$ MHz at modulation frequency below 1 MHz.}
\label{fig3}
\end{figure}

Obviously, for a practical application of the Rydberg atom based RF receiver, the SNR of the optically detected signal is a vital evaluation parameter.
We measured the optical signal at the center frequency $f_{1/2}$ versus the RF-field detuning $\Delta_{RF}$ at different modulation frequencies and the corresponding SNR values are shown in Fig. \ref{fig3}.
In the measurement, a wireless transmission of a $\sim$kHz sinusoidal signal of 2 V peak-to-peak amplitude over 10.22 GHz
carrier of -10 dBm power is tested. The demodulated signal is interrogated by a spectrum
analyzer (Keysight N9010B) with a resolution bandwidth of 100 Hz.
In Fig. \ref{fig3}(a), we can see that the SNR of the receiving signal performs well over the RF detuning ranging from -150 MHz to 150 MHz at three different modulation frequencies $\emph{f}_{mod}=1, 10 \rm{\ and \ } 100$ kHz. The SNR has a minimum value at RF detuning $\Delta_{RF}=50$ MHz.
It is due to the coupling of the neighboring state 59 $D_{3 / 2}$, exactly 50 MHz away from 59 $D_{5 / 2}$.
On the whole, the SNR has desirable values at  $\emph{f}_{mod}=10$ kHz compared to that at $\emph{f}_{mod}=1 \rm{\ and \ } 100$ kHz.
It can also be seen from Fig. \ref{fig3}(b) where the SNR values are measured for various modulation frequency at the resonant carrier. For current setup, the demodulated SNR arrives at the most desirable value SNR=70 dB at 10 kHz modulation. Of course, many factors can affect the obtained SNR, for example, the technical noise of the photodiode detector and the pre-amplifier, the coupling laser power and atomic density, etc.
However, fundamentally, the response of Rydberg atoms to an incident time-variant RF field is especially important since it determines the transfer rate in the communication. Physically, it is governed by the dynamic time for a steady EIT to rebuild-up, characterized by the inverse of its dephasing rate
\cite{66046}.

\begin{figure}[ptbh]
\centerline{\includegraphics[width=3.3in]{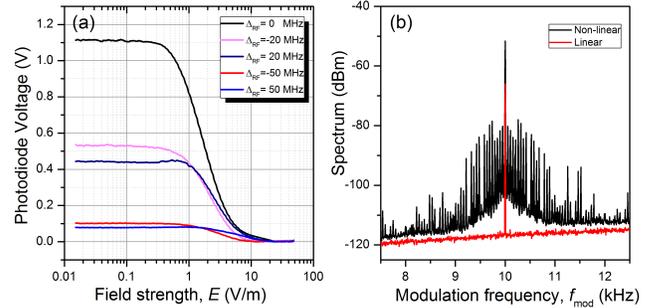}}
\caption{The response of the probe transmission to the applied RF-field over the RF-detuning range.
 The probe photodiode linearly responds to the RF-field in a wide dynamic range varying from weak field $E\sim1$ V/m to strong end $E\sim10$ V/m but has a maximum gain performance at zero detuning (a), which implies that a careful selection of linear response range will give a pure spectrum as indicated by the red line in (b).}
\label{fig4}
\end{figure}

Care should be taken to the nonlinearity of the photodiode gain responding to the RF-field amplitude over its detuning range. Wide response range and good linearity will lead to a better SNR performance. The measured results are presented in Fig. \ref{fig4}, where we can see that the photodiode has a good linear response for the RF-field varying from weak field of $\sim1$ V/m to strong end of $\sim10$ V/m, exhibiting a commendable performance in the applicable RF-field range. In this linear range, the best signal gain displays at zero detuning, $\Delta_{RF}=0$, hinting a nice SNR as shown in Fig. \ref{fig3}. It also indicates that choosing a moderate RF-power as working point is very vital in the reliable Rydberg atom based RF-field communication, which is qualitatively described in a testing shown in Fig. \ref{fig4}(b), where the black line represents the power spectrum  when nonlinear region is included in capturing a 10kHz signal over 10.22 GHz. It contains many nonlinear frequency modes. On the contrary, a careful selection for the working RF-field can single out all high frequency modes, implying a much better SNR in the practical communication. The red line is obtained at the linear response region of zero detuning at RF-power of -10 dBm. Similar analysis can be followed for communication at off-resonance RF carrier. With this linearity calibration, the power of the carrier and modulation signal can be optimized for practical long-distance communication by considering specific radio propagation models \cite{Seybold2005}.

\begin{figure}[ptbh]
\centerline{\includegraphics[width=3.3in]{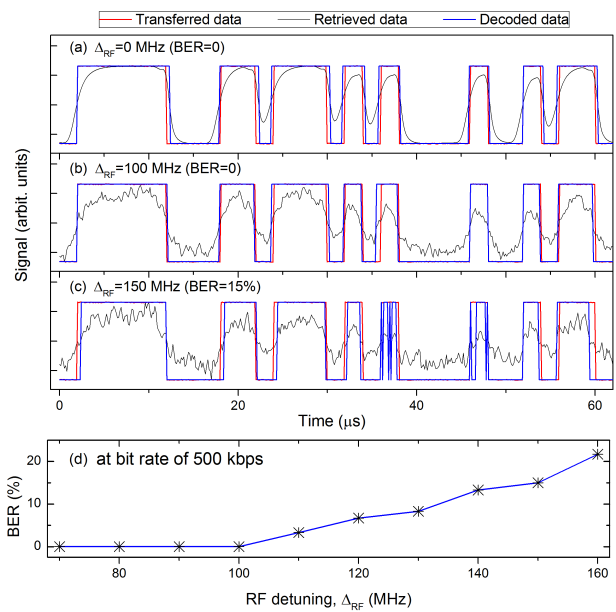}}
\caption{A digital communication tested using a benchmark of PRBS signal transferring at a bit rate of 500 kbps. Three typical waveform transfers are shown in (a), (b) and (c), respectively, corresponding to RF-detuning 0, 100 MHz and 150 MHz with respect to the on-resonant frequency (10.22 GHz). The BER determined by comparing the decoded and source digital signals rises up with RF-detuning, up to 20{\%} at $\Delta_{RF}=160$ MHz in (d).}
\label{fig5}
\end{figure}

Finally, we demonstrate a digital communication test using a PRBS signal transferred at different RF-carrier frequencies.
The PRBS signal recommended by the International Telecommunication Union (ITU), is commonly used in digital transmission testing \cite{bardell1987built}.
Its transfer over the Rydberg-based RF-receiver can provide a convincible support for its practical application prospect.
Figures \ref{fig5}(a)-\ref{fig5}(c) show three typical waveform transfers over different RF-detunings by the proposed off-resonance scheme for $\Delta_{RF}=0,100$ MHz and 150 MHz, respectively. The transfer bit rate is fixed at 500 kbps.
The red line waveforms stand for the source signal which are digitally encoded for wireless communication, and the black thin lines are the received optical signals using the Rydberg based RF-receiver. These optical signals are decoded back to the digital form as shown in blue. Comparing them with the source signal, we can obtain the bit error rate (BER), which is defined by a ratio of the number of bit errors to the total number of transferred bits during one cycle.
The waveform transfer performs well at RF-detunings $\Delta_{RF}=0 \rm {\ and \ }100$ MHz without any loss but a BER of 15$\%$ exists at $\Delta_{RF}=150$ MHz.

It implies the Rydberg based RF-receiving can provide a wideband carrier channel with high transfer fidelity.
It is proved by a further measurement of the transfer BER over the whole carrier frequency domain as shown in Fig. \ref{fig5}(d).
We can see that there isn't any loss of information over carrier RF-detuning from 0 to 100 MHz. Beyond $\Delta_{RF}=100$ MHz, the BER increases linearly with the detuning depth, with a BER of more than 20$\%$ at $\Delta_{RF}=160$ MHz.
Considering the negative detuning part, a bandwidth of 200 MHz with no loss can be achieved in our experiment.
Once the transfer bit rate increases to high values, for example, up to 1 Mbps, a larger BER is expected, a typical value of more than 30{\%} determined for 150 MHz off-resonance communication.

In summary, we have demonstrated broadband digital communication through high-sensitive RF-receiving based on the EIT of Rydberg atom in a room-temperature atomic vapor.
The RF-field is demodulated by detecting the intensity of a probe laser at the center frequency of RF-induced symmetric or asymmetric AT-splitting within an EIT window.
Various factors affecting the communication credibility within a wide continuous RF-detuning are investigated.
Besides the optimization of the coupling and probe laser powers \cite{Deb2018,17Meyer2018}, attention should also be paid to the choosing of a linear response of the probe density versus the RF field strength, frequency range of the modulation signal as well as the RF-detuning depth itself.
The digital communication testing using a standard PRBS signal transferring at a rate of 500 kbps shows a high fidelity for wireless communication within a tunable bandwidth of 200 MHz at carrier 10.22 GHz.
It is expected that the communication quality also relies on the selection of Rydberg states, as the effective RF-detuning depth varies with states.
This will inspire an attempt to employ many advanced communication techniques on the Rydberg atom-based RF receiver, such as the frequency-hopping spread spectrum (FHSS) and frequency division multiplexing (FDM) over the broadband RF-carrier.

\begin{acknowledgements}
The work is funded by the National Natural Science Foundation project (No.
91536110) and the National Key R{\&}D Program (No. 2016YFF0200104) of China.
\end{acknowledgements}

\bibliographystyle{apsrev4-1}
\bibliography{Ref}

\end{document}